\documentclass[aps, pra, reprint, superscriptaddress]{revtex4-1}
\usepackage{amssymb}
\usepackage{amsmath}
\usepackage{graphicx}
\usepackage{bm}
\usepackage{color}
\usepackage{lineno}

\begin{document}

\title{Static quantum dot on a potential hilltop \\
for generating and analyzing hot electrons in the quantum Hall regime}
\author{Ryo Oishi}
\author{Yuto Hongu}
\author{Tokuro Hata}
\author{Chaojing Lin}
\affiliation{Department of Physics, Tokyo Institute of Technology, 2-12-1 Ookayama,
Meguro, Tokyo, 152-8551, Japan.}
\author{Takafumi Akiho}
\author{Koji Muraki}
\affiliation{NTT Basic Research Laboratories, NTT Corporation, 3-1 Morinosato-Wakamiya,
Atsugi 243-0198, Japan.}
\author{Toshimasa Fujisawa}
\email{fujisawa@phys.titech.ac.jp}
\affiliation{Department of Physics, Tokyo Institute of Technology, 2-12-1 Ookayama,
Meguro, Tokyo, 152-8551, Japan.}
\date{\today }

\begin{abstract}
We propose and demonstrate a static quantum dot on a potential hilltop to
generate and analyze ballistic hot electrons along a quantum Hall edge
channel well above the chemical potential. High energy resolution associated
with discrete energy levels is attractive for studying hot-electron
dynamics. Particularly, the energy distribution function of hot electrons
weakly coupled to cold electrons is investigated to reveal spectral
diffusion with energy relaxation. The analysis allows us to estimate the
maximum energy exchange per scattering, which is an important parameter to
describe interacting electrons in the edge channel.
\end{abstract}

\maketitle

\section{Introduction}

Ballistic hot-electron transport in the quantum Hall regime is attractive
for studying electron dynamics and developing electronic quantum optics with
flying electrons, where highly non-equilibrium hot (high-energy) 
electrons are spatially and energetically separated from cold (low-energy) 
electrons near the 
chemical potential \cite{ReviewElecQO,ReviewElecQO2022,BookEzawa}. The crossed
electric and magnetic fields induce long ballistic transport over a
millimeter along a potential contour, when the scattering processes, such as
LO phonon emission and Coulomb interaction with cold electrons, are
well suppressed \cite%
{TaubertPRB2011,JohnsonPRL2018,AkiyamaAPL2019,EmaryPRB2019}. Such ballistic
transport can be seen by exciting electrons well above the chemical
potential typically by 30 -- 100 meV \cite{OtaPRB2019}. Unlike photons, a
strong anti-bunching correlation of two hot electrons has been identified as
a result of Coulomb repulsion, which can be used for designing non-linear
functionality \cite{UbbelohdeNatNano2023,FletcherNatNano2023}. Such
hot-electron dynamics are often studied with dynamic quantum dots (QDs), by
which a single hot electron is generated dynamically on-demand by applying
rapid voltage pulses \cite%
{Ubbelohde-NatNano2014,KataokaPRL2016,FletcherPRL2013,FletcherNatCom2019}.
In contrast, a single potential barrier is widely used as an energy
spectrometer, which has an energy resolution of only a few meV for typical
GaAs heterostructures \cite{OtaPRB2019,FletcherNatNano2023}. Quantum-dot
energy spectrometers with discrete energy levels are highly desirable for
higher energy resolution, which would allow us to distinguish coherent
transport from others and identify otherwise hidden inelastic processes.
However, typical QDs with multiple electrons work 
only at low energies close to the
chemical potential and cannot be used as a hot-electron spectrometer \cite%
{Altimiras-NatPhys10,leSueurPRL2010}.

Here, we demonstrate that a single static QD can be used for generating and
detecting hot electrons with a high energy resolution by preparing an empty QD
on a potential hilltop. Such a hilltop QD has Coulomb
charging energy and discrete energy spacing comparable to those of
conventional QDs for cold electrons. We find that hot electrons injected
from a point contact have an asymmetric energy distribution with a sharp
Fermi edge on the high-energy side and a broadened tail on the low-energy
side. When hot electrons are allowed to interact even weakly with cold
electrons, the hot electrons relax, accompanied by spectral diffusion, in
agreement with the numerical simulation of electron-electron scattering. The
relation between the relaxation and the spectral diffusion can be used to
extract the maximum energy exchange per scattering in the edge channel.

\section{Hilltop QD}

Figure 1(a) schematically shows how a QD is defined at the top of a
potential hill that separates two regions of 
different chemical potentials
(labeled B and R, which stand for base and reference, respectively). The
electrochemical potential $\varepsilon _{1}$ for the first electron in the
QD is located close to the chemical potential $\mu _{\mathrm{R}}$ (defined
to be $\mu _{\mathrm{R}}=0$) of the right region but much higher than that $%
\mu _{\mathrm{B}}=-eV_{\mathrm{B}}$ ($<0$) of the base region with large
bias voltage $V_{\mathrm{B}}$ ($>0$). When $\varepsilon _{1}$ is slightly
below $\mu _{\mathrm{R}}$, hot electrons with energy $\varepsilon _{1}$ can
be injected into the base with a narrow spectral width, as shown by the
energy distribution function $f_{\mathrm{h}}$ in the inset. The injected hot electrons are well isolated from cold electrons in the base. The hilltop QD
works as a narrow-band hot-electron injector under this condition.

\begin{figure}[tbp]
\begin{center}
\includegraphics[width = 3.3in]{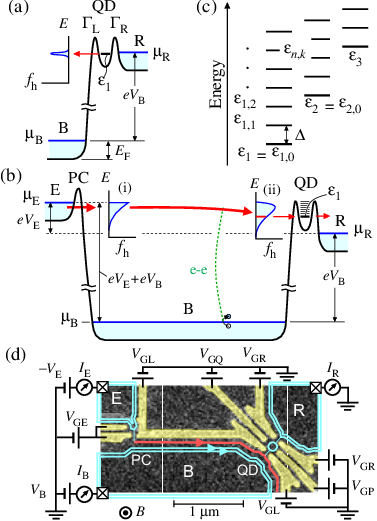}
\end{center}
\caption{(a) Energy diagram of a QD on a large potential hilltop between the
base (B) and the right reference (R) channels. A hot electron can be
injected from the electrochemical potential $\varepsilon _{1}$ for the first
electron in the QD. The inset shows the energy distribution function of the
injected hot electrons. (b) Energy diagram for analyzing hot electrons
emitted from the emitter (E) through a point contact (PC). The insets (i)
and (ii) show the energy distribution functions of hot electrons just after
the injection and after electron-electron scattering, respectively. (c)
Energy diagram of electrochemical potential $\varepsilon _{n,k}$ for the $k$%
-th $n$-electron state. Transport through $n$-electron ground state is
associated with $\varepsilon _{n}$ ($\equiv \varepsilon _{n,k}$). (d)
Schematic measurement setup with a false-color scanning electron micrograph
of a control sample. The cyan lines represent the cold edge channels, and
the red line shows the potential contour where ballistic hot electrons
travel. The polarity of the voltage on the emitter is intentionally reversed
to express the initial hot-electron energy as ($eV_{\mathrm{E}}+eV_{\mathrm{B%
}}$) in the base.}
\end{figure}

When $\varepsilon _{1}$ is made slightly above $\mu _{\mathrm{R}}$, as shown
in Fig. 1(b), hot electrons traveling to the right in the base can be
analyzed with the level $\varepsilon _{1}$ of the QD. To test this scheme,
hot electrons are injected through a point contact (PC) from the emitter
region (labeled E) with the chemical potential $\mu _{\mathrm{E}}=+eV_{%
\mathrm{E}}$ ($>0$) and voltage $-V_{\mathrm{E}}$ ($V_{\mathrm{E}}>0$). The
initial energy distribution function has a high-energy cutoff at $\mu _{%
\mathrm{E}}$ and a broad low-energy tail due to the weak energy dependence
of the tunneling probability, as shown in the inset (i). This sharp cutoff,
which should be determined by the base temperature $T_{\mathrm{B}}$ of the
system (the emitter region), can be used to evaluate the spectroscopic
resolution of the hilltop QD. If the hot electrons scatter with cold
electrons in the base, the energy distribution function at the analyzer
broadens, as shown in the inset (ii). Such electron-electron (e-e)
scattering can be investigated with a hilltop QD by changing the scattering
rate with $V_{\mathrm{B}}$ while keeping all other settings of the emitter
and spectrometer the same.

For these purposes, an empty hilltop QD has to be prepared with no
permanently occupied electrons. This is not a difficult task, because excess
electrons will escape from the QD to the base. Generally, $\varepsilon _{1}$
($\equiv \varepsilon _{1,0}$) for the first electron is the lowest one of
electrochemical potential $\varepsilon _{n,k}$ for $k$-th electronic state ($%
k=0,1,2,\cdots $ from the ground state) of $n$-electron QD ($n=1,2,3,\cdots $%
), as shown in Fig. 1(c). Even when the hot-electron energy distribution is
significantly broader than the level spacing $\Delta =\varepsilon
_{1,1}-\varepsilon _{1,0}$ of the QD, the spectroscopic analysis can be
performed dominantly with the lowest level $\varepsilon _{1}$, as shown in
the following experiment.

Such a hilltop QD was formed in a standard modulation-doped
AlGaAs/GaAs heterostructure 
with a two-dimensional electron system
(2DES) located 100 nm below the surface. 
As shown in Fig. 1(d), a PC and a QD can be formed with split gates 
in the same design as the device used in Ref. \cite{KonumaPRB2022}. 
All measurements were performed at the electron temperature of
$T_{\mathrm{B}}\simeq $ 100 mK (comparable to the lattice temperature). 
A magnetic field of 3.8 T was applied
perpendicular to the 2DES with electron density
of 1.7$\times $10$^{11}$ cm$^{-2}$ and low-temperature mobility of about 10$%
^{6}$ cm$^{2}$/Vs. The electrons occupy the Landau levels with the filling
factor $\nu =2$ in the bulk, and chiral edge channels are formed as shown by
the cyan lines, although integer filling is not required for this
experiment. At $\nu =2$, the Fermi level $E_{\mathrm{F}}$ of the edge state
measured from the lowest Landau level in the bulk should be about half the
cyclotron energy, 3 meV at 3.8 T. Under appropriate bias voltages of $V_{%
\mathrm{B}}$ (= 0 $-$ 100 mV) and $V_{\mathrm{E}}$ (= 0 $-$ 15 mV), gate
voltages $V_{\mathrm{G}m}$ on gate G$m$ with $m$ = E, L, P, Q, and R were
adjusted to form a hilltop QD and a PC separated by a distance of $L$ = 2 $%
\mu $m. 
Hot electrons injected from the emitter to the base will propagate along 
the potential contour (the red line).
The currents $I_{\mathrm{E}}$, $I_{\mathrm{B}}$, and $I_{\mathrm{R}}$
at the emitter, base, and reference regions, respectively are simultaneously
measured. See Appendix A and B for the quantum
Hall effect and potential profile of the device, respectively.

Standard Coulomb blockade oscillations with isolated current peaks in $I_{%
\mathrm{R}}$ are seen at small $V_{\mathrm{B}}$, as shown in Fig. 2(a) for $%
V_{\mathrm{B}}=0.1$ mV, where the two gate voltages $V_{\mathrm{GL}}$ and $%
V_{\mathrm{GR}}$ are swept at $V_{\mathrm{GP}}=V_{\mathrm{GQ}}=-300$ mV. The
current peak appears when the $n$-electron chemical potential $\varepsilon
_{n}$ ($\equiv \varepsilon _{n,0}$) is located between $\mu _{\mathrm{R}}$
and $\mu _{\mathrm{B}}$. When $V_{\mathrm{B}}$ is made slightly greater than
the charging energy $E_{\mathrm{C}}$ (about 2 meV) of the QD, the peaks are
broadened and partially overlapped, as shown in Fig. 2(b) for $V_{\mathrm{B}%
}=$ 3 mV. While the current is out of the range of the ammeter in the
hatched region, the Coulomb oscillations are visible as periodic kinks
(marked by the red lines) at $\varepsilon _{n}=\mu _{\mathrm{R}}$ along the
boundary of the conductive region. For both cases, the tunneling rates $%
\Gamma _{\mathrm{L}}$ and $\Gamma _{\mathrm{R}}$ of the left and right
barriers are primarily controlled by the gate voltages $V_{\mathrm{GL}}$ and $%
V_{\mathrm{GR}}$, respectively, and thus the current vanishes if either $%
\Gamma _{\mathrm{L}}$ or $\Gamma _{\mathrm{R}}$ becomes lower than the
measurement limit ($\Gamma _{\mathrm{\lim }}\sim $ 1 MHz for 160 fA).
Therefore, the last peak in Fig. 2(a) and the last kink in Fig. 2(b) seen at
the lower left end of the conductive region should be determined by this
measurement limit, not by emptying the QD.

\begin{figure}[tbp]
\begin{center}
\includegraphics[width = 3.3in]{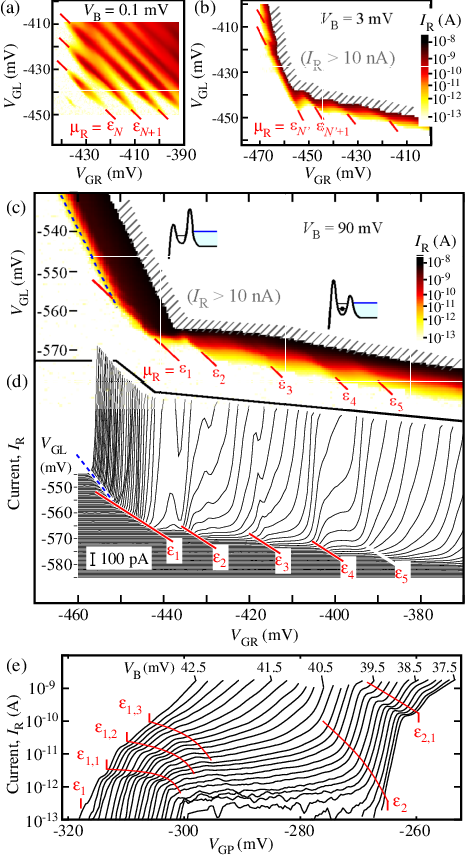}
\end{center}
\caption{(a) Standard Coulomb blockade oscillations in the current $I_{%
\mathrm{R}}$ measured at low bias $V_{\mathrm{B}}=$ 0.1 mV. (b) Partially
overlapped Coulomb oscillations at $V_{\mathrm{B}}=$ 3 mV greater than the
charging energy of about 2 meV. (c and d) Coulomb oscillations of the
hilltop QD at large $V_{\mathrm{B}}=$ 90 mV in the color scale plot (c) and
the linear line plot (d) of the same data. Each line in (d) is offset for clarity. (e) Current steps in $I_{\mathrm{R%
}}$ as a function of the plunger gate voltage $V_{\mathrm{GP}}$ for various $%
V_{\mathrm{B}}$. The peak, kink, and step structures in (a-e) are associated
with electrochemical potentials $\varepsilon _{n}$ ($\equiv \varepsilon
_{n,0}$) and $\varepsilon _{n,k}$ of the QD.}
\end{figure}

The condition to form the hilltop QD for the first electron can be found by
sweeping $V_{\mathrm{GL}}$ and $V_{\mathrm{GR}}$ under a large $V_{\mathrm{B}%
}$, as shown in the color plot of Fig. 2(c) and the line plot of Fig. 2(d)
for the same data set taken at $V_{\mathrm{B}}$ = 90 mV. 
Because the left barrier is electrostatically influenced by the
large $V_{\mathrm{B}}$, large negative $V_{\mathrm{GL}} \simeq$ -570 mV is needed to pinch 
off the channel.
The conductive
region is terminated by the left smooth boundary and periodic kinks along
the bottom boundary. The period of the kinks is comparable to those in Fig.
2(a) and (b), which suggests that the kinks are related to the Coulomb
oscillations at $\varepsilon _{n}=\mu _{\mathrm{R}}$ of the hilltop QD. The
current profiles in Fig. 2(d) show clear Coulomb staircases around the kinks 
\cite{BookIhn}. The staircases appear only along the bottom boundary, where
the left barrier is higher ($\Gamma _{\mathrm{L}}\ll \Gamma _{\mathrm{R}}$)
and electrons can accumulate in the QD [the energy diagram shown in the
right inset to Fig. 2(c)]. In contrast, the QD should be fully emptied in
the opposite limit, $\Gamma _{\mathrm{L}}\gg \Gamma _{\mathrm{R}}$, along
the left boundary (left inset), which explains the absence of kinks along
the left boundary. Namely, the left boundary marked by the blue dashed line
should be determined by the measurable limit for the right barrier ($\Gamma
_{\mathrm{R}}\sim \Gamma _{\mathrm{\lim }}$) with the QD empty. Under this
condition, the QD states do not contribute to the transport. Therefore, the
last kink at the lower left corner of the conductive region should be
attributed to the transport through the first electron state of the QD at $%
\varepsilon _{1}=\mu _{\mathrm{R}}$ with comparable tunneling rates $\Gamma
_{\mathrm{L}}\sim \Gamma _{\mathrm{R}}$. One can use this condition for
generating and detecting hot electrons in the base.

Discrete energy levels in the hilltop QD can be resolved in the fine sweep
of plunger gate voltage $V_{\mathrm{GP}}$, as shown in Fig. 2(e). Here, the
current profile is obtained at various $V_{\mathrm{B}}$ values and fixed $V_{%
\mathrm{GL}}=-0.5$ V and $V_{\mathrm{GR}}=-0.453$ V. Between the large
current onsets for the first electron at $\mu _{\mathrm{R}}=\varepsilon _{1}$
and the second electron at $\varepsilon _{2}$, step-wise features marked by
the red lines are observed. These current steps are attributed to the
excited states at $\varepsilon _{1,k}$. By using the lever arm factor $%
\alpha =$ 0.055 for $V_{\mathrm{GP}}$ estimated from the following
measurement, we obtained the onsite charging energy $E_{\mathrm{C}%
}=\varepsilon _{2}-\varepsilon _{1}=$ 1.8 meV and the level spacing $\Delta
=\varepsilon _{1,1}-\varepsilon _{1}=$ 0.2 meV. These values are comparable
to those for a conventional QD with a similar device structure \cite%
{KonumaPRB2022,SuzukiCommPhys2023,WashioPRB2016,Itoh-PRL2018}.

For our electrostatically-defined QD, the overall current decreases with decreasing $V_{\mathrm{B}}$, which can be
understood by considering that the left barrier height is increased with
decreasing $V_{\mathrm{B}}$  (See Appendix C 
for the relation to the Coulomb diamond characteristics). 
It should be noted that the current steps
associated with the discrete levels are not clearly seen for small $V_{%
\mathrm{B}} <$  39 mV. This can be understood by considering the
competition between energy relaxation within and escape from the QD. If $%
\Gamma _{\mathrm{L}}$ is much greater than the relaxation rate $\Gamma _{%
\mathrm{rel}}$, electrons in the QD excited levels will escape before
relaxing to the ground state, where the current reflects the state-dependent 
$\Gamma _{\mathrm{L}}$. For example, the data at $V_{\mathrm{B}}>$ 42 mV
suggests that $\Gamma _{\mathrm{L}}$ for the first excited state is about 10
times greater than $\Gamma _{\mathrm{L}}$ for the ground state. For $\Gamma
_{\mathrm{L}}\ll \Gamma _{\mathrm{rel}}$, on the other hand, electrons in
the excited state will relax to the ground state before escaping \cite%
{FujisawaJPhysCM2003}. This explains why the current remains constant in the
range of $\varepsilon _{1}<\mu _{\mathrm{R}}<\varepsilon _{2}$ for the data
at $V_{\mathrm{B}}<$ 39 mV. In this range, the hilltop QD can be used as a
hot-electron injector with a tunable kinetic energy $eV_{\mathrm{B}}-\left(
\mu _{\mathrm{R}}-\varepsilon _{1}\right) $ in the base. The injected
electrons should have a narrow spectral width comparable to the transition
width from the zero current to the constant current (about 20 $\mu $eV for
the present data). Ideally, the width can be reduced to the lifetime
broadening $\hbar \Gamma _{\mathrm{L}}$ ($\simeq $0.004 $\mu $eV for 1 pA).
Such a monochromatic hot-electron source is attractive for studying
hot-electron dynamics.

\section{Energy spectroscopy for hot electrons}

\subsection{QD spectrometer}

We apply this hilltop QD as a spectrometer to investigate ballistic
hot-electrons injected from a point contact (PC), as shown in Fig. 1(b). The
hot-electron energy distribution function $f_{\mathrm{h}}(\varepsilon)$,
which describes the occupation probability $0 \le f_{\mathrm{h}} \le 1$ 
for a state at energy $\varepsilon$, 
can be measured
with the hilltop QD at $\varepsilon _{1}>0$, where only one-electron states
with $\varepsilon _{1,k}$ contribute to the transport. If the energy
distribution is narrower than the level spacing $\Delta $ of the QD, the
current $I_{\mathrm{R}}$ is proportional to $\Gamma ^{\left(
1,k\right) }f_{\mathrm{h}}\left( \varepsilon _{1,k}\right) $ for the
transport through the $k$-th state with the overall tunneling rate 
$\Gamma ^{\left( 1,k\right) }$. $f_{\mathrm{h}}\left( \varepsilon \right) $ can be
measured directly from the $\varepsilon _{1,k}$ dependence of $I_{\mathrm{R}%
} $. In our experiment using a PC as the injector, the distribution is much
wider than $\Delta $. Therefore, the current $I_{\mathrm{R}}\propto
\sum_{k}\Gamma ^{\left( 1,k\right) }f_{\mathrm{h}}\left(
\varepsilon _{1,k}\right) $ is contributed by a multiple of one-electron
states $k$. If $\varepsilon _{1,k}$ is equally spaced from the minimum $%
\varepsilon _{1}$, and if $\Gamma ^{\left( 1,k\right) }$ is
independent of the states, the current can be approximated to be $I_{\mathrm{%
R}}\propto \int_{\varepsilon _{1}}^{\infty }f_{\mathrm{h}}\left( \varepsilon
\right) d\varepsilon $. In this case, the distribution function $f_{\mathrm{h%
}}\left( \varepsilon \right) \propto -\frac{d}{d\varepsilon _{1}}I_{\mathrm{R%
}}$ can be obtained from the derivative of $I_{\mathrm{R}}$ with respect to $%
\varepsilon _{1}$. The validity of the scheme is
discussed with experimental data, as shown below.

In the experiment, the emitter voltage $-V_{\mathrm{E}}$ ($V_{\mathrm{E}}=$
0 -- 15 mV) is applied relative to the reference electrode, and thus the
injected hot electron has maximum kinetic energy $E_{\mathrm{EB}}=eV_{%
\mathrm{E}}+eV_{\mathrm{B}}$ in the base. The hot electrons are injected
with constant current $I_{\mathrm{E}}=$ 1 nA by tuning the emitter gate
voltage $V_{\mathrm{GE}}$ with a feedback program after setting all
experimental conditions ($V_{\mathrm{E}}$, $V_{\mathrm{GL}}$, $V_{\mathrm{GR}%
}$, $V_{\mathrm{GP}}$, etc.). 
For this small $I_{\mathrm{E}}$, we assume that spin-up electrons 
in the emitter enter the lowest Landau level in the base.
Because the hot-electron distribution function
should be small ($f_{\mathrm{h}}\lesssim $ 0.01) for this $I_{\mathrm{E}}$,
we adjusted the gate voltages for the QD to yield large tunneling rates $\Gamma _{%
\mathrm{L}}$ and $\Gamma _{\mathrm{R}}$ to obtain sufficiently large
detector current in $I_{\mathrm{R}}$. Under the hot electron injection, the
QD current $I_{\mathrm{R}}$ is measured as a function of $V_{\mathrm{GL}}$
and $V_{\mathrm{GR}}$, as shown in the inset to Fig. 3(a). In addition to
the positive $I_{\mathrm{R}}$ due to electron escape from the QD to the
base, which is accompanied by Coulomb oscillations with kinks (the red
lines), we observe in some regions negative $I_{\mathrm{R}}$ (the blue
region), which is a manifestation of hot electrons entering the QD. Large
negative $I_{\mathrm{R}}$ appears near the last kink at $\mu _{\mathrm{R}%
}\lesssim \varepsilon _{1}$ for the first electron. This is the situation
where the hilltop QD can probe the hot electrons. In contrast, small
negative $I_{\mathrm{R}}$ is seen along the bottom boundary (the blue region
at $V_{\mathrm{GR}}>-360$ mV), where multiple electrons contribute to the
transport. No negative current is seen along the left boundary ($V_{\mathrm{%
GL}}>-460$ mV), where the well-defined QD may not have formed. These
characteristics ensure that the hot electrons are primarily detected with
one-electron QD states in the large-negative-$I_{\mathrm{R}}$ region ($V_{%
\mathrm{GR}}<-380$ mV and $V_{\mathrm{GL}}<-460$ mV). 
 See Appendix D for further discussions on the hot-electron detection.
The following
measurements were performed in the middle of the large-negative-$I_{\mathrm{R%
}}$ region marked by the circle, where the two barriers are expected to have
comparable tunneling rates ($\Gamma _{\mathrm{L}}\simeq \Gamma _{\mathrm{R}}$%
).

\begin{figure}[tbp]
\begin{center}
\includegraphics[width = 3.3in]{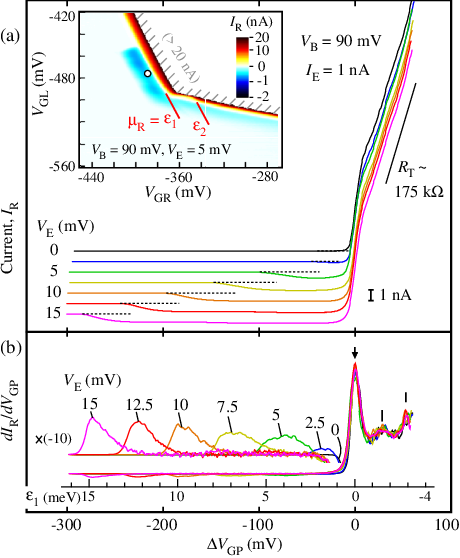}
\end{center}
\caption{(a) Current $I_{\mathrm{R}}$ as a function of the plunger gate
voltage $\Delta V_{\mathrm{GP}}$ taken at several $V_{\mathrm{E}}$ values.
Each line is offset for clarity. The dashed lines represent the zero level ($%
I_{\mathrm{R}}=0$). The inset shows the color scale plot of $I_{\mathrm{R}}$
obtained by a two-dimensional sweep of $V_{\mathrm{GL}}$ and $V_{\mathrm{GR}%
} $. The QD condition for the main panel is marked by the circle in the
inset. (b) $dI_{\mathrm{R}}/dV_{\mathrm{GP}}$ for the data in (a).
Sign-reversed magnified plots [$\times \left( -10\right) $] are also shown.
The horizontal axis $\Delta V_{\mathrm{GP}}$ is measured from the largest
peak in $dI_{\mathrm{R}}/dV_{\mathrm{GP}}$ and is converted to $\varepsilon
_{1}=-\alpha \Delta V_{\mathrm{GP}}$ as shown in the inset scale.}
\end{figure}

The energy spectroscopy was performed by measuring $I_{\mathrm{R}}$ as a
function of $V_{\mathrm{GP}}$, as shown in the main plot of Fig. 3(a). Its
derivative $dI_{\mathrm{R}}/dV_{\mathrm{GP}}$ is shown in Fig. 3(b). The
largest current step in positive $I_{\mathrm{R}}$ (the largest peak marked
by the arrow in $dI_{\mathrm{R}}/dV_{\mathrm{GP}}$) is attributed to the
resonance at $\mu _{\mathrm{R}}=\varepsilon _{1}$, and thus the horizontal
axis $\Delta V_{\mathrm{GP}}$ was defined by the shift of $V_{\mathrm{GP}}$
from the resonant condition. The energy scale for $\varepsilon _{1}=-\alpha
\Delta V_{\mathrm{GP}}$ is also shown in Fig. 3(b) by assuming a linear
shift of $\varepsilon _{1}$ with $\alpha =$ 0.055. Before analyzing the hot
electron spectroscopy, we examine the QD characteristics seen in the
positive-$I_{\mathrm{R}}$ region. In Fig. 3(a), $I_{\mathrm{R}}$ increases
almost linearly with $\varepsilon _{1}$, which suggests a tunneling
resistance of $R_{\mathrm{T}}$ = 175 k$\Omega $ including the series
resistance in the setup. Under this high tunneling conductance $R_{\mathrm{T}%
}^{-1}$, discrete levels of the QD are unresolved even in the $dI_{\mathrm{R}%
}/dV_{\mathrm{GP}}$ traces in Fig. 3(b). The peaks in $dI_{\mathrm{R}}/dV_{%
\mathrm{GP}}$ at $\Delta V_{\mathrm{GP}}\simeq $ 0, 30 and 52 meV (marked by
the vertical bars) seems to be associated with the charging energy (about
1.5 meV) of the QD. The large peak at $\Delta V_{\mathrm{GP}}=0$ could be
enhanced by the Fermi edge singularity \cite%
{FermiEdge-EnsslinG2017,FermiEdge-Theory}. If we neglect the peak
structures, $dI_{\mathrm{R}}/dV_{\mathrm{GP}}$ at $\Delta V_{\mathrm{GP}}>0$
is almost constant in agreement with the Fermi distribution function 
and the constant density of edge states. This supports that
the proposed scheme with $f_{\mathrm{h}}\left(\varepsilon \right)%
\propto \frac{d}{d\varepsilon _{1}}I_{\mathrm{R}}$ works for this 
highly-conductive QD. Whereas we did not resolve discrete levels here,
the overall tunneling rate $\Gamma ^{\left(1,k\right)}$ can be 
regarded as a constant in this energy regime.
The scheme should work even better for hot electrons, as many-body effects
should be absent for sparse electrons ($f_{\mathrm{h}}\ll 1$).

For the data at $\Delta V_{\mathrm{GP}}<0$ in Fig. 3(a), the hot electrons
yield a negative current $I_{\mathrm{R}}$ $\simeq $ $-$0.8 nA comparable in
amplitude to $I_{\mathrm{E}}=$ 1 nA, which suggests that the QD is quite
transparent above $\varepsilon _{1}$. The negative current disappears at
large negative $\Delta V_{\mathrm{GP}}$. The vanishing point [the peak
position in the magnified $-dI_{\mathrm{R}}/dV_{\mathrm{GP}}$ plot of Fig.
3(b)] shifts almost linearly with increasing $V_{\mathrm{E}}$, which
supports the linear energy shift of $\varepsilon _{1}=-\alpha \Delta V_{%
\mathrm{GP}}$ for this energy range. A small deviation from linearity (not
clear in this plot) will be discussed later in connection with relaxation.
Therefore, $-dI_{\mathrm{R}}/dV_{\mathrm{GP}}$ as a function of $\Delta V_{%
\mathrm{GP}}$ can be viewed as a distribution
function $f_{\mathrm{h}}\left( \varepsilon _{1}\right) $. The distribution
function is asymmetrically broadened particularly at large $V_{\mathrm{E}}$
= 15 mV.

The broadening is clearly seen in the logarithmic plot of $-dI_{\mathrm{R}%
}/dV_{\mathrm{GP}}$ as a function of $\varepsilon _{1}$ in Fig. 4(a) for the
data taken at different $V_{\mathrm{B}}$ values. Here, the relative energy
between the emitter and the detector is fixed at $V_{\mathrm{E}}$ = 11 meV
 to keep the same emitter-detector configuration.
The spectral change with $V_{\mathrm{B}}$ 
can be explained by the electron-electron scattering. 
As illustrated in Fig. 4(b), the scattering is reduced by increasing 
the physical distance between hot and cold electrons at 
larger $V_{\mathrm{B}}$ \cite{OtaPRB2019,SuzukiCommPhys2023}.
Importantly, all spectra are terminated at $\varepsilon _{1}\simeq \mu _{%
\mathrm{E}}=$ 11 meV. The observation of the high-energy cutoff at $\mu _{%
\mathrm{E}}$ ensures that a fraction of electrons transport ballistically
over a distance $L=$ 2 $\mu $m. For $\varepsilon _{1}>\mu _{\mathrm{E}}$, $%
-dI_{\mathrm{R}}/dV_{\mathrm{GP}}$ decays exponentially as $e^{-\varepsilon
_{1}/\rho }$ with $\rho \simeq $ 0.2 meV at $V_{\mathrm{B}}=$ 90 mV.
Considering that the thermal energy ($k_{\mathrm{B}}T_{\mathrm{B}}\simeq $
10 $\mu $eV)\ of the emitter is much smaller than $\rho $, $\rho $ should be
determined by the resolution of the highly transparent QD.

\begin{figure}[t]
\begin{center}
\includegraphics[width = 3.1in]{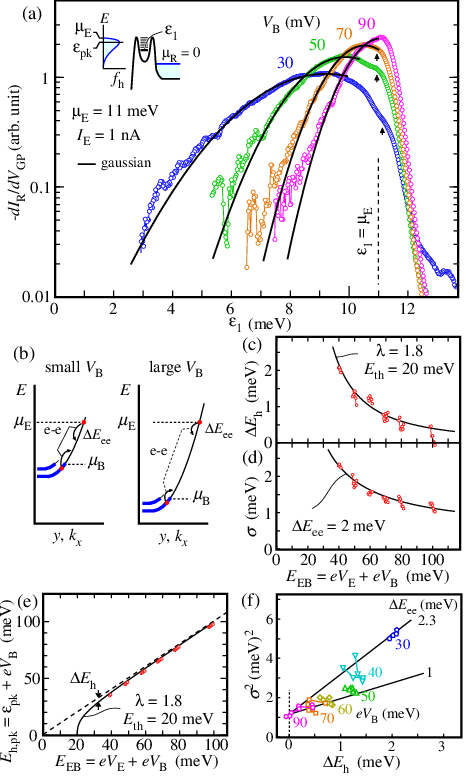}
\end{center}
\caption{(a) Measured $-dI_{\mathrm{R}}/dV_{\mathrm{GP}}$ (the circles) taken
at several $V_{\mathrm{B}}$ values as a function of $\varepsilon _{1}$
converted from $V_{\mathrm{GP}}$. The profile can be regarded as the
distribution function $f_{\mathrm{h}}\left( \varepsilon _{1}\right) $. The
dashed line for $\varepsilon _{1}=\mu _{\mathrm{E}}$ (= 11 meV) represents
the high-energy cutoff by the emitter chemical potential. The low-energy
profile on the left side of the peak was fitted with a Gaussian function
(the black solid lines) with the peak position $\varepsilon _{\mathrm{pk}}$
and the standard deviation $\sigma $. The inset shows the energy diagram of
the measurement.
(b) Energy ($E$) - momentum ($k_{x}$) dispersion relations of the edge
channels at small and large $V_{\mathrm{B}}$. The e-e scattering with 
maximum energy exchange $\Delta E_{\mathrm{ee}}$ is shown.
(c) The energy loss $\Delta E_{\mathrm{h}}=eV_{\mathrm{E}%
}-\varepsilon _{\mathrm{pk}}$, (d) $\sigma $, and (e) the peak energy $E_{%
\mathrm{h.pk}}=\varepsilon _{\mathrm{pk}}+eV_{\mathrm{B}}$ as a function of
the energy bias $E_{\mathrm{EB}}=eV_{\mathrm{E}}+eV_{\mathrm{B}}$. The solid
lines in (c-e) are based on the empirical model with parameters $\lambda =$
1.8, $E_{\mathrm{th}}=$ 20 meV, and $\Delta E_{\mathrm{ee}}=$ 2 meV (see
text). (f) $\sigma ^{2}$ as a function of $\Delta E_{\mathrm{h}}$ to
estimate $\Delta E_{\mathrm{ee}}$ from the slope.}
\end{figure}

In contrast, the spectra at $\varepsilon _{1}<\mu _{\mathrm{E}}$ are
significantly broadened. The spectrum taken at $V_{\mathrm{B}}$ = 90 mV
shows a peak position close to $\mu _{\mathrm{E}}$, which suggests
negligible e-e scattering. With decreasing $V_{\mathrm{B}}$, the peak is
shifted to the low-energy side. As all spectra are asymmetric around the
peak, the low-energy side of the peak is fitted by a Gaussian function, as
shown by the black curves, from which the peak position $\varepsilon _{%
\mathrm{pk}}$ and the standard deviation $\sigma $ are obtained. The energy
loss $\Delta E_{\mathrm{h}}=eV_{\mathrm{E}}-\varepsilon _{\mathrm{pk}}$ and
the broadening $\sigma $ are obtained for various conditions, as summarized
in Fig. 4(c) and (d) with fine tuning of $V_{\mathrm{E}}$ = 5 -- 11 mV
(connected by the red lines) and coarse tuning of 
$V_{\mathrm{B}}$ = 40 -- 90 meV in the step of 10 mV. 
They are plotted as a function of the
excitation energy $E_{\mathrm{EB}}=eV_{\mathrm{E}}+eV_{\mathrm{B}}$ in the
base. Generally, both $\Delta E_{\mathrm{h}}$ and $\sigma $ decrease with
increasing $E_{\mathrm{EB}}$, which can be understood by the reduction of
e-e scattering, because the hot electrons are spatially separated from the
cold electrons in the base. While $\Delta E_{\mathrm{h}}$ approaches to zero
at $E_{\mathrm{EB}}\gtrsim $ 100 meV, finite $\sigma \sim $ 1 meV remains.
This residual $\sigma $ should be related to the energy dependence of the
tunneling probability of the emitter PC.

Previously, the relation between the initial and final energy after
traveling a finite distance has been investigated by using point-contact
spectroscopy \cite{OtaPRB2019}. This relation can be tested using the
hilltop QD with higher energy resolution. We plot the initial energy $E_{%
\mathrm{EB}}$ and final energy $E_{\mathrm{h},\mathrm{pk}}=\varepsilon _{%
\mathrm{pk}}+eV_{\mathrm{B}}$ at $L$ = 2 $\mu $m, as shown in Fig. 4(e). The
lever arm factor $\alpha $ = 0.0547 was determined to make this plot so that 
$E_{\mathrm{h},\mathrm{pk}}$ calculated with $\varepsilon _{\mathrm{pk}%
}=-\alpha e\Delta V_{\mathrm{GP,pk}}$ from the peak position $\Delta V_{%
\mathrm{GP,pk}}$ in $\Delta V_{\mathrm{GP}}$ asymptotically approaches the
ballistic condition (the dashed line) at large $E_{\mathrm{EB}}$. Whereas
the deviation from the ballistic condition is not clear in this plot, finite
deviation $\Delta E_{\mathrm{h}}$ is identified with the high-resolution QD
spectrometer, as shown in Fig. 4(c).

Generally, hot electrons can also be scattered by emitting 
LO phonons and tunneling into other Landau levels. 
Such unwanted processes should induce additional signals well separated from 
the ballistic signal by the LO phonon energy (about 36 meV) and 
the cyclotron energy (about 6 meV) \cite{OtaPRB2019}. 
Because we focused on the narrow energy range around the ballistic condition, 
the present analysis is made only for the hot electrons that 
have not experienced these processes. Since we obtain
a large signal ($|I_{\mathrm{R}}/I_{\mathrm{E}}| \simeq 0.8$), 
the unwanted processes should play minor roles in the present analysis.

\subsection{e-e scattering}

The measurement of distribution functions allows us to characterize the e-e
scattering process in detail. The scattering can be characterized by two
parameters; the maximum energy exchange $\Delta E_{\mathrm{ee}}$ per
scattering due to the non-momentum-conserving scattering and the effective
interaction strength $\gamma $ per length [see Eq. (9) in Ref. \cite%
{LundePRB-EE}]. Particularly, $\Delta E_{\mathrm{ee}}$ is the important
parameter that distinguishes the single-particle excitations ($E>\Delta E_{%
\mathrm{ee}}$) and many-body excitations ($E<\Delta E_{\mathrm{ee}}$) for
excitation energy $E$. We have previously estimated $\Delta E_{\mathrm{ee}}$
by measuring the electron-hole excitation near the Fermi energy \cite%
{SuzukiCommPhys2023}. Here, we estimate $\Delta E_{\mathrm{ee}}$ from the
hot-electron distribution function and provide a complementary data set.

We employ the single-particle scattering approach shown in Ref. \cite%
{LundePRB2010-EE,LundePRB-EE}. Whereas $\Delta E_{\mathrm{ee}}\ll k_{\mathrm{%
B}}T_{\mathrm{B}}$ was assumed there, we used $\Delta E_{\mathrm{ee}}\gg k_{%
\mathrm{B}}T_{\mathrm{B}}$ in the following analysis, because $\Delta E_{%
\mathrm{ee}}>$ 0.5 meV (much greater than $k_{\mathrm{B}}T_{\mathrm{B}%
}\simeq $ 10 $\mu $eV) was suggested from the electron-hole excitation \cite%
{SuzukiCommPhys2023}. By assuming $f_{\mathrm{h}}\left( E\right) \ll 1$, the
master equation for spatial ($x$) evolution of $f_{\mathrm{h}}\left(
E;x\right) $ reads%
\begin{align}
& \frac{\partial }{\partial x}f_{\mathrm{h}}\left( E;x\right) =
\label{Eqmaster} \\
& \gamma \int_{0}^{\infty }d\epsilon e^{-\left( \frac{\epsilon }{\Delta E_{%
\mathrm{ee}}}\right) ^{2}}D_{\mathrm{c}}\left( \epsilon \right) \left[ f_{%
\mathrm{h}}\left( E+\epsilon ;x\right) -f_{\mathrm{h}}\left( E;x\right) %
\right] ,  \notag
\end{align}%
where the scattering from energy $E+\epsilon $ to $E$ with the energy loss $%
\epsilon $ ($>0$) is suppressed by the factor $e^{-\left( \frac{\epsilon }{%
\Delta E_{\mathrm{ee}}}\right) ^{2}}$ with parameter $\Delta E_{\mathrm{ee}}$%
. Here, $D_{\mathrm{c}}\left( \epsilon \right) =\int_{-\infty }^{\infty
}dEf_{\mathrm{c}}\left( E-\epsilon \right) \left[ 1-f_{\mathrm{c}}\left(
E\right) \right] $ is the density of electron-hole excitation of cold
electrons with the distribution function $f_{\mathrm{c}}\left( E\right) $
and is approximated to be $D_{\mathrm{c}}\left( \epsilon \right) =\epsilon $
for $\epsilon \geq 0$ and $0$ for $\epsilon \leq 0$ at low temperature ($T_{%
\mathrm{B}}=0$).

In the present case, the initial distribution function is expected to be $f_{%
\mathrm{h}}\left( E;0\right) =f_{0}\exp \left( \frac{E-\mu _{\mathrm{E}}}{w}%
\right) $ with an exponential decay for $E\leq \mu _{\mathrm{E}}$ and $f_{%
\mathrm{h}}\left( E;0\right) =0$ with a sharp cut off for $E>\mu _{\mathrm{E}%
}$. $f_{0}$ ($\ll 1$) is a constant that depends on the emitter current $I_{%
\mathrm{E}}=\frac{e}{h}\int_{-\infty }^{\infty }f_{\mathrm{h}}\left(
E;0\right) dE$. The spatial evolution $f_{\mathrm{h}}\left( E;x\right) $ can
be obtained by integrating Eq. (\ref{Eqmaster}), as shown in Fig. 5(a)
calculated with $w=$ 1 meV and $\Delta E_{\mathrm{ee}}=$ 1 meV. The initial
function is transformed toward a Gaussian-like function. Interestingly, the
sharp peak at $E=\mu _{\mathrm{E}}$ remains visible even if the ballistic
signal $f_{\mathrm{h}}\left( \mu _{\mathrm{E}};x\right) $ is heavily
attenuated by the scattering, because scattering with small energy loss ($%
\epsilon \simeq 0$) is suppressed by the factor $D_{\mathrm{c}}\left(
\epsilon \right) $. This ballistic signal can be smeared out if the detector
energy resolution is not sufficiently high. We consider an energy window
function $g\left( E\right) =\frac{1}{\sqrt{2\pi }\rho }\exp \left( -\frac{%
E^{2}}{2\rho ^{2}}\right) $ to describe the energy resolution of width $\rho 
$, and the measurement yields a convolution $\tilde{f}_{\mathrm{h}}\left(
E\right) =\left( f_{\mathrm{h}}\ast g\right) \left( E\right) $ of $f_{%
\mathrm{h}}\left( E\right) $ and $g\left( E\right) $. The convoluted $\tilde{%
f}_{\mathrm{h}}\left( E;x\right) $ with $\rho =$ 0.2 meV is shown in Fig.
5(b), which qualitatively reproduces the relaxation and broadening of the
experimental data in Fig. 4(a). 
 Interestingly, the signature of the ballistic transport is seen as
the small bumps marked by the vertical bars at $E = \mu _{\mathrm{E}}$ 
in Fig. 5(b) and probably by the arrows at $\varepsilon _{1} = \mu _{\mathrm{E}}$ 
in Fig. 4(a). The similarity implies that 10 -- 60 \% of detected electrons 
have transported ballistically over $L =$ 2 $\mu$m for 
$V _{\mathrm{B}} =$ 30 -- 90 meV.

For comparison, the same simulation is
performed with $\Delta E_{\mathrm{ee}}=$ 2 meV in Fig. 5(c) showing more
broadening, and $\Delta E_{\mathrm{ee}}=$ 0.5 meV in Fig. 5(d) showing less
broadening. Here, the range of $\gamma x$ values for each $\Delta E_{\mathrm{%
ee}}$ is chosen to yield comparable peak shifts by using the scaling [$%
\gamma x\propto \left( \Delta E_{\mathrm{ee}}\right) ^{-3}$]. The simulation
indicates that $\gamma $ and $\Delta E_{\mathrm{ee}}$ can be determined from
the peak shift and the broadening. Crude comparison with the data suggests $%
\Delta E_{\mathrm{ee}}\sim $ 1 meV in our sample.

\begin{figure}[tbp]
\begin{center}
\includegraphics[width = 3.3in]{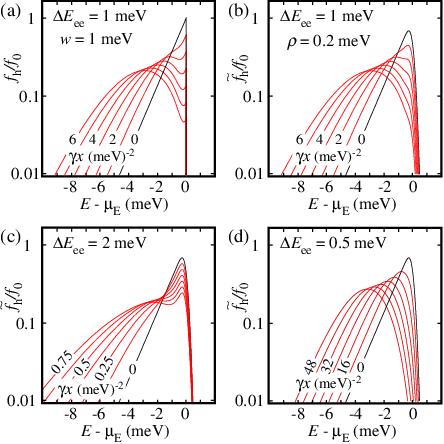}
\end{center}
\caption{(a) Calculated evolution of the energy distribution function $f_{%
\mathrm{h}}\left( E\right) $ of hot electrons injected from an emitter with
chemical potential $\mu _{\mathrm{E}}$. (b-d) Expected evolution of
convoluted distribution function $\tilde{f}_{\mathrm{h}}\left( E\right)
=\left( f_{\mathrm{h}}\ast g\right) \left( E\right) $ with an energy window
function $g\left( E\right) $ with width $\rho $ = 0.2 meV. The maximum
energy exchange per scattering is $\Delta E_{\mathrm{ee}}=$ 1 meV in (a) and
(b), 2 meV in (c), and 0.5 meV in (d). The black trace in each panel shows
the initial function with $w=$ 1 meV.}
\end{figure}

Since our measurement was performed at a fixed distance $L$ = 2 $\mu $m, the
spectroscopic change with $V_{\mathrm{B}}$ in Fig. 4(a) suggests variations
of $\gamma $ and/or $\Delta E_{\mathrm{ee}}$ with $V_{\mathrm{B}}$. To
evaluate the parameters, we expand $f_{\mathrm{h}}\left( E+\epsilon \right)
\simeq f_{\mathrm{h}}\left( E\right) +\epsilon \frac{\partial }{\partial E}%
f_{\mathrm{h}}\left( E\right) +\frac{1}{2}\epsilon ^{2}\frac{\partial ^{2}}{%
\partial E^{2}}f_{\mathrm{h}}\left( E\right) $ up to the second order of $E$
to derive a Fokker-Planck equation of the form%
\begin{equation}
\frac{\partial }{\partial x}f_{\mathrm{h}}\left( E;x\right) =\eta \left[ 
\frac{\partial }{\partial E}f_{\mathrm{h}}\left( E\right) +\frac{\Delta E_{%
\mathrm{ee}}}{\sqrt{\pi }}\frac{\partial ^{2}}{\partial E^{2}}f_{\mathrm{h}%
}\left( E\right) \right]  \label{EqFokkerPlanck}
\end{equation}%
from Eq. (\ref{Eqmaster}). Here, $\eta \equiv \frac{\sqrt{\pi }}{4}\gamma
\left( \Delta E_{\mathrm{ee}}\right) ^{3}$ describes the energy-relaxation
velocity $\eta =\frac{d}{dx}\left\langle E\right\rangle $. By assuming a
Gaussian form of the solution%
\begin{equation}
f_{\mathrm{h}}\left( E;x\right) =\frac{f_{\mathrm{0}}}{\sqrt{2\pi }\sigma }%
\exp \left[ -\frac{\left( E-\left\langle E\right\rangle \right) ^{2}}{%
2\sigma ^{2}}\right] ,
\end{equation}
the average energy $\left\langle E\left( x\right) \right\rangle
=\left\langle E_{0}\right\rangle -\eta x$ and the standard deviation $\sigma
\left( x\right) =\sqrt{\sigma _{0}^{2}+\frac{2}{\sqrt{\pi }}\eta \Delta E_{%
\mathrm{ee}}x}$ changes with $x$ from the initial values $\left\langle
E_{0}\right\rangle $ and $\sigma _{0}$. Therefore, $\eta =\Delta E_{\mathrm{h%
}}/L$ (= 0 - 1 meV/$\mu $m) can be obtained from the energy loss $\Delta E_{%
\mathrm{h}}$ in Fig. 4(c). $\Delta E_{\mathrm{ee}}$ can be estimated from
the combined relation 
\begin{equation}
\sigma ^{2}=\sigma _{0}^{2}+\frac{2}{\sqrt{\pi }}\Delta E_{\mathrm{ee}%
}\Delta E_{\mathrm{h}},  \label{EqSigma2}
\end{equation}%
i.e., the slope of the $\sigma ^{2}-\Delta E_{\mathrm{h}}$ plot in Fig.
4(f). $\Delta E_{\mathrm{ee}}$ has a weak energy dependence; about 1 meV at $%
E>$ 50 meV and 2.3 meV at $E\sim $ 30 meV, in good agreement with our previous
work (about 1 meV at $E=$ 50 meV and 2 meV at $E=$ 30 meV in Fig. 9 of Ref 
\cite{SuzukiCommPhys2023}). The coincidence with different measurement
schemes justifies the proposed scheme and ensures the obtained parameter.

The energy dependence of $\Delta E_{\mathrm{ee}}$ can be understood with the
concave edge potential profile, which can be regarded as the one-dimensional
energy $E$ - momentum $k_{x}$ relation, as shown in Fig. 4(b).
For non-linear dispersion, e-e scattering must be accompanied by a finite
change in the total momentum, which is allowed in the presence of random impurity
potential. The maximum momentum change $\Delta k_{x}$, which is related to
the correlation length of the impurity potential, induces the maximum energy
exchange $\Delta E_{\mathrm{ee}}=\hbar \left\vert v_{\mathrm{c}}^{-1}-v_{%
\mathrm{h}}^{-1}\right\vert ^{-1}\Delta k_{x}$. Here, $v_{\mathrm{h}}$ and $%
v_{\mathrm{c}}$ are the drift velocities $v=\frac{1}{\hbar }\frac{d}{dk_{x}}%
E $ for hot and cold electrons, respectively. This qualitatively explains
why $\Delta E_{\mathrm{ee}}$ increases with decreasing the hot-electron
energy.

The initial-energy ($E_{\mathrm{EB}}$) dependence of $E_{\mathrm{h},\mathrm{%
pk}}$ in Fig. 4(e) as well as $\Delta E_{\mathrm{h}}$ in Fig. 4(c) can be
understood by phenomenological energy dependence of $\eta \propto
E^{-\lambda }$, where the power $\lambda $ (= 0.5 $\sim $ 4) depends on the
details of the edge potential profile, the screening effect from the metal
gate, and the velocity of the cold electrons in the base \cite{OtaPRB2019}. $%
\lambda =2$ is suggested for non-interacting cold electrons and negligible
screening effect. This empirical power law is convenient as it provides an
analytical solution of $\frac{d}{dx}\left\langle E\right\rangle =\eta $. For
initial average energy $\left\langle E_{0}\right\rangle $, $\left\langle
E\left( L\right) \right\rangle $ at $x=L$ is given by%
\begin{equation}
\left\langle E\left( L\right) \right\rangle =\left[ \left\langle
E_{0}\right\rangle ^{1+\lambda }-E_{\mathrm{th}}^{1+\lambda }\right] ^{\frac{%
1}{1+\lambda }}  \label{EqTrajectory}
\end{equation}%
with parameters $\lambda $ and $E_{\mathrm{th}}$. By regarding $E_{\mathrm{EB%
}}$ as $\left\langle E_{0}\right\rangle $ and $E_{\mathrm{h},\mathrm{pk}}$
as $\left\langle E\left( L\right) \right\rangle $ in Fig. 4(e), the
experimental data can be reproduced with Eq. (\ref{EqTrajectory}), as shown
by the solid line with $\lambda =$ 1.8 and $E_{\mathrm{th}}=$ 20 meV. These
values are comparable to those in the previous study \cite%
{OtaPRB2019,SuzukiCommPhys2023}. Moreover, the $\sigma $ data in Fig. 4(d)
can be reproduced with Eq. (\ref{EqSigma2}), where a constant $\Delta E_{%
\mathrm{ee}}=$ 2 meV was used for simplicity as its energy dependence is not
precisely determined in this experiment. In this way, the relaxation
dynamics of hot electrons is successfully investigated with the hilltop QD.

\section{Summary}

In summary, we developed a hilltop QD to generate and analyze hot electrons
whose energy is well above the chemical potential of cold electrons. Taking
advantage of discrete energy levels in the QD, the scheme can be used as a
high-energy-resolution emitter and spectrometer. We applied this technique
to resolve the asymmetric energy distribution function of hot electrons
injected from the emitter. The distribution is significantly broadened by
Coulomb interaction with cold electrons. The analysis with a master equation
characterizes the energy loss and the broadening with some parameters. The
developed hilltop QD can be used to study ballistic hot-electron transport
in various ways. Whereas the QD spectroscopy was demonstrated with high
tunneling conductance in this experiment, the resolution can be improved
further by making the tunneling rates smaller. As suggested in this
experiment, the hilltop QD can be used as a hot-electron emitter with a
narrow spectral width to launch a spatially broadened wave packet. If
required, the hilltop QD can be driven with a voltage pulse to launch a hot
electron on demand and to detect hot electrons in a time-resolved fashion 
\cite{Kamata-NatNano2014,Hashisaka-NatPhys2017,LinNatComm2021}. The present
scheme should be useful in studying ballistic hot electron transport.

\bigskip

\textbf{Acknowledgements}

This study was supported by the Grants-in-Aid for Scientific Research
(KAKENHI JP19H05603, JP23K17302, JP24K00552, and JP24H00827) and "Advanced
Research Infrastructure for Materials and Nanotechnology in Japan (ARIM)" of
the Ministry of Education, Culture, Sports, Science and Technology (MEXT).

\bigskip

\appendix
\renewcommand{\figurename}{FIG. S}
\setcounter{figure}{0}

\section{Quantum Hall effect}

The quantum Hall effect of the device used in this study is shown in the Hall
resistance $R_{xy}$ and the longitudinal resistance $R_{xx}$ of Fig. S1,
where all gate voltages were set to zero. 
Whereas the ohmic contacts and the conductive region of the device are not 
designed for a standard Hall bar, 
clear quantized Hall conductance in $R_{xy}$ and zero
longitudinal resistance in $R_{xx}$ are resolved in some magnetic-field regions
including $B$ = 3.8 T where the measurement in the main paper was
performed.

\begin{figure}[bp]
\begin{center}
\includegraphics[width = 3.3in]{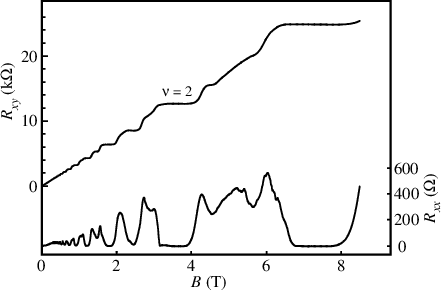}
\end{center}
\caption{The Hall resistance $R_{xy}$ and the longitudinal resistance $R_{xx}$ 
as a function of the magnetic field $B$.}
\end{figure}

\section{Potential profile}

It may be useful to see the potential profile for a better understanding of the device 
characteristics. Figure S2 shows a schematic potential profile of the device
for analyzing hot electrons injected from a point contact (PC) with
a hilltop quantum dot (QD). This potential profile was obtained by calculating the
electrostatic potential at 100 nm below the surface, where the two-dimensional 
electron system (2DES) should exist, under a given surface potential 
profile \cite{DaviesJAP1995}. 
The surface potentials in the gated and ungated regions are chosen 
for illustration purposes. For simplicity, we neglected the potential induced by
the charges of the 2DES and donors. While the profile should be regarded as a schematic,
one can see how the PC and QD can be formed at the tops of the potential hill from the base side. 

\begin{figure}[bp]
\begin{center}
\includegraphics[width = 3.3in]{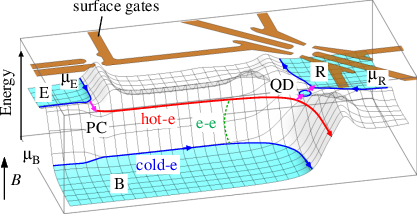}
\end{center}
\caption{Schematic potential profile of the device and the surface gate
pattern. }
\end{figure}

With the grounded reference electrode labeled R with chemical potential 
$\mu _{\mathrm{R}} = 0$, slightly larger chemical potential 
$\mu _{\mathrm{E}} = eV_{\mathrm{E}}$ ($>0$) is set to the emitter labeled E, 
and large negative potential $\mu _{\mathrm{B}}=-eV_{\mathrm{B}}$ ($<0$) is 
applied to the base labeled B. The 2DESs in the emitter, base, and reference 
regions are shown by cyan color and corresponding edge
channels (only for spin-up) are shown by the blue lines. 
Hot electrons injected from the PC into the base have significantly large 
kinetic energy as compared to cold electrons in the edge channels of the base. 
The hot electrons propagate along the equipotential line (the red line) based on the
drift motion under the crossed electric and magnetic fields. The hot
electrons may experience electron-electron (e-e) scattering during the
propagation. The scattering strength depends on the physical distance
between hot and cold electrons and can be reduced by
increasing $V_{\mathrm{B}}$, as shown in the main paper. The energy
distribution function of hot electrons is investigated by using the hilltop
QD.

As discussed in the main text, e-e scattering is sensitive to the edge 
potential primarily determined by the base potential $\mu _{\mathrm{B}}$ 
and the gate voltage $V_{\mathrm{GL}}$.
$V_{\mathrm{GL}}$ has to be sufficiently negative to define the transport 
path of hot electrons but not too much negative to suppress unwanted processes 
like LO phonon scattering and inter-Landau-level scattering, because these 
processes become efficient if the edge potential is steeper.

\section{Coulomb diamond}

While Coulomb diamond characteristics are widely investigated for
conventional QDs, we do not observe clear Coulomb diamond for the hilltop
QD. Figure S3 shows schematic Coulomb diamond characteristics when the gate
voltage $V_{\mathrm{GP}}$ and the bias voltage $V_{\mathrm{B}}$ are swept 
in the wide range. 
The Coulomb blockade regions for electron number $n=$ 1, 2, ... appear as 
diamond patterns in this diagram,
and the maximum bias voltage within the Coulomb blockade condition is
determined by the charging energy of the QD. For the demonstrated
hot-electron spectroscopy, we utilize the semi-infinitely wide Coulomb blockade region of
the empty QD ($n=0$), where the bias voltage can exceed the charging energy.

\begin{figure}[tbp]
\begin{center}
\includegraphics[width = 3.1in]{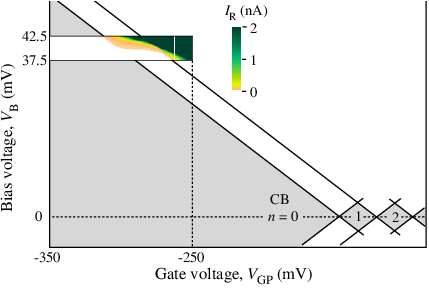}
\end{center}
\caption{Schematic charging diagram of the QD in the $V_{\mathrm{GP}}-V_{%
\mathrm{B}}$ plane. CB regions are shown in gray.
A color plot of $I_{\mathrm{R}}$ [the same data as in Fig.
2(e)] is attached.}
\end{figure}

The current $I_{\mathrm{R}}$ shown in Fig. 2(e) is attached to Fig. S3 
as a color-scale plot, where the onsets for $n=$ 1 and 2 transport regions 
are adjusted to the schematic charging diagram. 
Because the voltage $V_{\mathrm{B}}$ on
the base strongly influences the left tunneling barrier of the QD, the
current level changes drastically with $V_{\mathrm{B}}$. Therefore, the onsets 
for $n=$ 1 and 2 are visible only in the limited regions. 
It is not an easy task to observe the entire Coulomb diamond characteristics 
in a single $V_{\mathrm{B}}-V_{\mathrm{GP}}$ scan. 
In this study, every time $V_{\mathrm{B}}$ is changed, 
we swept $V_{\mathrm{GL}}$ and $V_{\mathrm{GR}}$ to find the condition 
for the last kink, as shown in Fig. 2(c) for example at $V_{\mathrm{B}}=$ 90 mV.

\section{Hot-electron detection}

\begin{figure}[bp]
\begin{center}
\includegraphics[width = 3.1in]{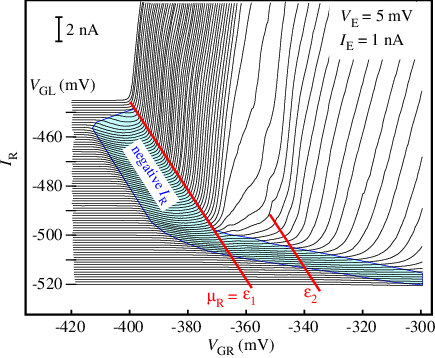}
\end{center}
\caption{Line plots of $I_{\mathrm{R}}$ as a funnction of $V_{\mathrm{GR}}$
for different $V_{\mathrm{GL}}$. Each line is offset for clarity.}
\end{figure}

A color plot of the current $I_{\mathrm{R}}$ shown in the inset to Fig. 3(a)
is shown as line plots in Fig. S4, where the current onsets for $n=$ 1 ($\mu
_{\mathrm{R}}=\varepsilon _{1}$) and 2 ($\mu _{\mathrm{R}}=\varepsilon _{2}$%
) are highlighted by the red lines. 
Other onsets for $n\geq $ 3 are smeared out because we adjusted 
the $n=1$ QD highly conductive (still in the tunneling regime) to obtain sufficient 
detector current, and thus the dot for $n>2$ may not be formed in the tunneling regime. 
Because the negative $I_{\mathrm{R}}$ is seen in the vicinity of the
last onset along $\mu _{\mathrm{R}}=\varepsilon _{1}$, the QD should play an
essential role in the detection. 

However, several features cannot be explained with the simple model. 
For instance, the maximum detector current
is very high ($I_{\mathrm{R}} =$ -0.8 nA close to $I_{\mathrm{E}} =$
1 nA in amplitude) in the wide region near the last onset 
$\mu _{\mathrm{R}} \leq \varepsilon _{1}$ in Fig. S4.
We need further studies to investigate how the hilltop QD efficiently 
carries hot electrons.


\end{document}